\documentstyle[11pt,blois2002]{article}
\def\Journal#1#2#3#4{{#1} {\bf #2}, #3 (#4)}

\def\NPB{{\em Nucl. Phys.} B}
\def\PLB{{\em Phys. Lett.}  B}
\def\PRL{\em Phys. Rev. Lett.}
\def\PRD{{\em Phys. Rev.} D} 
 
\def\CMP{\em Comm. Math. Phys.}
\def\JPA{{\em J. Phys} A}
\def\PTP{\em Prog. Theo. Phys.}
\def\SJNP{\em Sov. Jour. Nucl. Phys.}
\begin{document}
\title{Thoughts on CP Violation}
\author{R. D. Peccei}
\address{Department of Physics and Astronomy, UCLA, Los Angeles, 
CA 90095-1547}
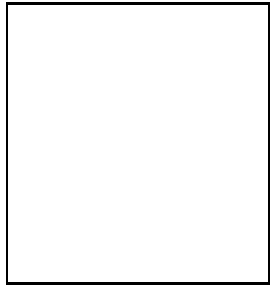
\maketitle
\begin{abstract}

The necessary complex structures needed for CP violation are not easy to generate and, indeed, CP can be a symmetry in higher dimensional theories. In 4-dimensions CP violation argues for the existence of a scalar sector and, in its simplest manifestation, leads to the CKM model. Further CP violating phases, from an extended scalar sector, are constrained by the requirement of having no FCNC. Although new CP violating phases are expected in SUSY extensions of the SM, the good agreement of data with the CKM model only provides bounds on these phases.  Baryogenesis via leptogenesis requires the presence of CP violating phases in the neutrino sector. However, the CP phases likely to be eventually meausured experimentally are not necessarily directly connected to the ones which drive the leptogenesis.

\end{abstract}

\section{Violating CP Is Not Easy}

For CP (or T) to be violated one must have complex structures in the
theory.\cite{Pec5}  This is easy to understand since under CP, effectively,
operators are replaced by their Hermitian adjoints.  For instance, for the 
charged $SU(2)$ gauge field $W_+^\mu$ under CP
\begin{equation}
W_+^\mu({\bf x},t) \to \eta(\mu) W_-^\mu({\bf -x},t),
\end{equation}
where $\eta(0) = -1$ and $\eta(i) = +1$.  Thus, schematically, under CP
an operator $O({\bf x},t)$ is replaced by:
\begin{equation}
O({\bf x},t) \to O^\dagger(-{\bf x},t).
\end{equation}
Because Lagrangians must be Hermitian, a term in the Lagrangian containing
the operator $O$ has the structure
\begin{equation}
{\cal{L}} = aO + a^\star O^\dagger~,
\end{equation}
where $a$ is a $c$-number.  It is clear from the above that ${\cal{L}}$ is
CP-invariant if $a^\star = a$.  So, without some complexity there will be
no CP violation.

Having some complex structures in the theory, however, by itself may not
suffice to give CP violation.  A well known example is the two-generation
Standard Model (SM) in which the phases in the complex Yukawa couplings can
always be rotated away.  There are also cases in which CP-violating terms
appear in the theory without an apparent complex structure.  The most well
known example is the famous $\theta F^{\mu\nu}\tilde F_{\mu\nu}$ term of QCD
which gives rises to the strong CP problem.  However, this term can be 
traced to the complex superposition of states which enters in the proper
gauge theory vacuum.\cite{JRCDG}  A more challenging problem, to which I return below, is why this
allowed CP-violating term is apparently absent to a high degree of accuracy.

Although we observe that CP is violated experimentally, in certain contexts
it is much more natural for CP to be conserved.  In fact, as noted by Dine,
Leigh, and MacIntire \cite{DLM} and Choi, Kaplan and Nelson, \cite{CKN} based on an original observation of
Strominger and Witten,\cite{SW} in 10-dimensional heterotic string theory
CP can be identified with the product of a Lorentz transformation in the 10-dimensional spacetime times a gauge transformation.  In this theory
fermions and gauge fields are in the adjoint representation of $E_8$, which is
real, and CP acts as an inversion in a 6-dimensional compact space of the
10-dimensional theory.  Thus, if our 4-dimensional world should originate from
such a theory, the observed CP-violating effects have their origin as a result
of the 10d $\to$ 4d compactification.  In these kinds of theories, in 
principle, one may be able to compute the resulting 4-dimensional CP-violating
phases from the underlying geometry.\cite{Abel}  I will return to this
interesting possibility at the end of this paper.

Even in four dimensions, violating CP is not easy.  In particular, a theory
involving only fermions and gauge fields is CP-conserving, up to $\theta$-terms.
Such theories have real coupling constants $g_i$, since the corresponding
gauge fields $A_i^\mu$ transform according to the adjoint representation of
the respective groups.  The topological nature of the non-Abelian gauge
theory vacuum, in the Standard Model allows for the presence of two 
CP-violating $\theta$-terms:
\begin{equation}
{\cal{L}} = \theta_W \frac{\alpha_2}{8\pi} W_a^{\mu\nu}
\tilde W_{a\mu\nu} + \theta_S \frac{\alpha_3}{8\pi} F_a^{\mu\nu}
\tilde F_{a\mu\nu}.
\end{equation}
However, because the electroweak theory is chiral, the $\theta_W$ term
can be rotated away.\cite{Rub}  Furthermore, as mentioned above, the
$\theta_S$ term is severely restricted, since no evidence has been found yet
for a neutron electric dipole moment.  The present bound on this moment
[${\rm edm} < 6.3\times 10^{-26}~{\rm ecm}]~$\cite{PDG} requires$~$\cite{bound}
$\theta_S < 10^{-10}$. This is the strong CP problem.

The strong CP problem is still unresolved,\cite{PecF} with four possibilities
being bruited  about:
\begin{description}
\item{i)} It could be that $\theta_S$ accidentally happens to be small, just like other
parameters in the SM---like the ratio $m_e/m_t$.  However, this is hardly
satisfactory as an explanation!
\item{ii)} The strong interactions also have a chiral symmetry, connected to the
vanishing of the $u$-quark mass, which effectively allow $\theta_S$ to be rotated away.  However, a careful current algebra analysis
of the hadronic spectrum$~$\cite{L} appear to argue against this possibility.
\item{iii)} The SM Lagrangian is augmented by an additional chiral global
symmetry $U(1)_{\rm PQ}$,\cite{PQ}  which forces $\theta_S \to 0$ dynamically. In this case the parameter $\theta_S$ is effectively
replaced by a dynamical axion field $[\theta_S\to a(x)/f_a]$.  However, axions have not been seen, and the scale $f_a$ of $U(1)_{\rm PQ}$ breaking is severely
limited.  Nevertheless, axions remain a tantalizing and beautiful candidate
for the Universe's dark matter.\cite{Sikivie}
\item{iv)} CP is a spontaneously broken symmetry and $\theta_S$ is a
calculably small parameter.  Although models exist where this is realized,\cite{BN} in general these models run into difficulties either with
cosmology (see below) or cannot reproduce the structure of the
observed CP violation at low energy.
\end{description}

\noindent
At any rate, whatever the reason is for $\theta_S < 10^{-10}$, it is
clear that this term by itself cannot be at the origin of the observed CP
violation in K and B physics.  These CP-violating phenomena are connected
to flavor-changing transitions, while the $\theta_S F\tilde F$ term is
flavor diagonal.

To account for the observed CP-violating phenomena, if there are {\bf no}
elementary scalar fields, it is necessary to imagine the formation of
CP-violating fermion condensates.  These, most likely, would need
to involve some other fermion fields (techni-fermions) rather than the 
ordinary quarks and leptons.  However, the formation of complex CP-violating
condensates $\langle\bar TT\rangle\sim e^{i~\delta_{TC}}\Lambda^3_{TC}$, with
$\Lambda_{TC}\sim G_F^{-1/2}$, is very problematic.\cite{PecN}  The origin
of this problem was pointed out long ago by Zeldovich, Kobzarev and Okun.\cite{KOZ}  They showed that, if CP is a spontaneously broken 
symmetry,
domains of different CP will form in the Universe, separated by walls whose slow
dissipation with temperature is a cosmological catastrophe.  Indeed, the energy
density in the domain walls only decreases linearly with temperature,
$\rho\sim \sigma T$.  If $\sigma$ is of order $\sigma\sim G_F^{-3/2}$ this 
energy density greatly exceeds the closure density of the Universe and the
model makes no sense.  One can countenance spontaneous violation of CP only
if the scale where this violation occurs is above the scale where inflation
takes place, since then one can inflate the domains away.  However, this cannot
happen in models where the fermion condensates must also serve to break
$SU(2)\times U(1)$.\cite{PecN}

\section{CP Violation and the Scalar Sector}

In view of the above discussion, it seems very natural to assume that the
experimentally observed CP violation is due to the presence of a scalar
sector in the theory.  Indeed, personally I think that the existence of CP violation
at low energy is as compelling evidence for a Higgs field as are the
precision electroweak tests which suggest the presence of a light Higgs
boson, $M_H < 204~{\rm GeV}$ at 95\% C. L..\cite{PDG}  In fact, as discussed in
detail by Buras$~$\cite{Buras} at this Rencontre, all data in both K and B
decays are perfectly consistent with the CKM paradigm,\cite{CKM} where all the observed CP violation originates from the complex Yukawa couplings of the
Higgs field with the quarks.  As is well known, with three generations of
quarks, this model has just one physical phase $\delta_{\rm CKM}$.  All
data appear consistent with this phase, in the standard convention, being
rather large:  $\delta_{\rm CKM} \sim (59 \pm 13)^{\circ}$.\cite{Kleinknecht}

It is clearly very important to look for deviations from the CKM paradigm, but
both data and theory at this moment do not allow us to make any such
pronouncement.  I will return to this important point later on.  However,
first I want to discuss theoretically whether it might be possible to identify
sources of possible {\bf flavor conserving} CP-violating effects coming from
the pure Higgs sector itself.  As we shall see, these effects are not easy
to find if one takes into account the structure of what we know about the
weak interactions!

The SM, in which only one Higgs doublet is introduced, is very special.  In this
case, the required Hermiticity of the Higgs potential makes all parameters in the potential real:
\begin{equation}
V = \mu^2\phi^\dagger\phi + \lambda(\phi^\dagger\phi)^2~.
\end{equation}
Thus, there are no additional CP-violating phases in the SM besides $\delta_{\rm CKM}$. However,
if there is more than one Higgs doublet, the Higgs potential in general can
have some possible CP-violating phases.  But, even in this case, there are
constraints.  It is useful to illustrate the nature of these constraints in the
case of having two Higgs doublets
\begin{equation}
\chi = \left( \begin{array}{c}
\chi^+ \\  \chi^o 
\end{array} \right) ~;  \qquad
\phi = \left( \begin{array}{c}
\phi^o \\ \phi^-
\end{array} \right).
\end{equation}

The most general Higgs potential in the 2-doublet model can be written as the sum of three terms, reflecting specific additional symmetries besides
$SU(2)\times U(1)$.  To understand this structure, it is useful to recall
that under the weak hypercharge $U(1)$ the Higgs doublets in Eq. (6) transform as
\begin{equation}
\chi\to e^{i\xi/2}\chi~; \qquad \phi\to e^{-i\xi/2}\phi~.
\end{equation}
There is an additional Abelian symmetry, $U(1)_{\rm PQ}$, which one can contemplate
for these fields, which rotates them in the same way:
\begin{equation}
\chi\to e^{i\alpha}\chi~; \qquad \phi\to e^{i\alpha}\phi.
\end{equation}
Such a symmetry allows for a chiral transformation of the quarks, thereby
setting $\theta_S\to 0$ dynamically.\cite{PQ}  Finally, one can also consider
possible discrete symmetries for the $\chi$ and $\phi$ fields.  In particular,
to avoid flavor changing neutral currents (FCNC)$~$\cite{GW} one can
consider a discrete symmetry $D$ which allows $\phi$ only to couple to
$u_R$ and $\chi$ only to couple to $d_R$.  Under $D$:
\begin{equation}
\chi\to -\chi~; \quad
\phi\to\phi~; \quad d_R\to -d_R~; \quad
u_R\to u_R.
\end{equation}

The full Higgs potential is the sum of three terms: $V = V_1 + V_2 + V_3$. The first term, $V_1$, is 
$SU(2)\times U(1)\times U(1)_{\rm PQ} \times D$ invariant; $V_2$ is
$SU(2)\times U(1)\times D$ invariant; and $V_3$ is only invariant
under the electroweak group.  In detail, one has{\footnote{Here C is a charge conjugation matrix.}}
\begin{eqnarray}
V_1 &=& \mu_1^2\chi^\dagger\chi + \mu_2^2\phi^\dagger\phi +
\lambda_1(\chi^\dagger\chi)^2 + \lambda_2(\phi^\dagger\phi)^2 \nonumber \\ 
& &\mbox{}+\lambda_3(\phi^\dagger\chi)(\chi^\dagger\phi) + \lambda_4
(\chi^\dagger\chi)(\phi^\dagger\phi) \\
\noalign{\vspace{4pt}}
V_2 &=& \lambda_5 e^{i\delta_5}(\chi^TC\phi)^2 + \lambda_5
e^{-i\delta_5}(\chi^TC\phi)^{2\dagger} \\
\noalign{\vspace{4pt}}
V_3 &=& \mu_{12}^2 e^{i\delta_{12}}(\chi^TC\phi) + \mu_{12}^2
e^{-i\delta_{12}}(\chi^TC\phi)^\dagger \nonumber \\
& &\mbox{} +\left[\lambda_6 e^{i\delta_6}(\chi^TC\phi) + \lambda_6e^{-i\delta_6}
(\chi^TC\phi)^\dagger\right]\chi^\dagger\chi \nonumber \\
& &\mbox{} +\left[\lambda_7e^{i\delta_7}(\chi^TC\phi)
+ \lambda_7 e^{-i\delta_7}(\chi^TC\phi)^\dagger\right]
\phi^\dagger\phi~.
\end{eqnarray}
One sees therefore, that if one asks that $V$ be just $SU(1)\times U(1)$
invariant, the full Higgs potential contains 4 phases: 
$\delta_5;~\delta_{12};~\delta_6$; and $\delta_7$.  If, on the other hand,
one asks that $V$ be also $U(1)_{\rm PQ}$ invariant (so that $V=V_1$)
all of the possible Higgs sector CP violating phases disappear.

If only the discrete symmetry $D$ is present (so that $V=V_1+V_2$), one
additional Higgs sector phase, $\delta_5$, appears in the potential.
However, as Branco, Lavoura and Silva$~$\cite{BLS} note in their nice book
on CP violation, this phase gives no physical CP-violating effects.
The phase $\delta_5$ in this case is correlated directly with the phase
of the Higgs VEVs $\theta$:
\begin{equation}
\langle\chi^o\rangle = v_\chi~; \qquad
\langle\phi^o\rangle = v_\phi e^{i\theta}~.
\end{equation}
Minimization of the potential $V = V_1 + V_2$ requires that
${\rm sin}(\delta_5 + 2\theta) = 0$.  It is easy to check that all
CP-violating phenomena, like for example the coupling of the axial Higgs
field $A$ to $H^+H^-$, are proportional to the phase combination
$\delta_5+2\theta$, which vanishes:
\begin{equation}
g_{AH^+H^-} \sim \sin (\delta_5 + 2\theta) = 0
\end{equation}
Thus, remarkably, even in the case of having 2 Higgs doublets, the requirement
that there be {\bf no FCNC} (i.e. that $D$ be a good symmetry) prevents the
appearance of any other CP-violating phases, besides the CKM phase
$\delta_{\rm CKM}$.

There are a number of corollaries to this result.  For instance, in 
axion models$~$ \cite{invisible} where $U(1)_{\rm PQ}$ is broken at a scale
$f_a \gg v\sim G_F^{-1/2}$, no additional CP-violation ensues in the 
Higgs sector.  In such models the spontaneous breaking of $U(1)_{\rm PQ}$
is effected by a complex singlet field $\sigma$, with VEV $\langle\sigma\rangle
\simeq f_a$.
The $U(1)_{\rm PQ}$ invariant potential
\begin{equation}
V_a = \kappa e^{i\delta_a}\sigma^2(\chi^TC\phi) + \kappa e^{-i\delta_a}
[\sigma^2(\chi^TC\phi)]^\dagger,
\end{equation}
with $\kappa f_a^2 \equiv \mu_a^2\sim v^2$, generates an additional complex
term to the Higgs potential, beyond $V_1$.  However, also in this case, the
phase $\delta_a$ (just like $\delta_5$  in the previous discussion) is
locked to the phase $\theta$ of the doublet Higgs VEV, and no physical
CP-violating effects ensue.

It is, of course, possible to get Higgs sector CP-violating effects by complicating the theory further.  The simplest case involves introducing an
additional real singlet field $\eta$, which is also odd under $D~
(\eta\to -\eta)$  The total potential $V$ now is $V = V_1 + V_2 + V_4$,
where $V_4$ contains an additional phase $\delta_4$:
\begin{eqnarray}
V_4 &=& \kappa^\prime(\eta^2-f^2)^2 + \mu^2_4 e^{i\delta_4}\eta(\chi^TC\phi)
\nonumber \\
& & \mbox{} + \mu^2_4 e^{-i\delta_4}\eta(\chi^TC\phi)^\dagger.
\end{eqnarray}
Because $\eta$ acquires a VEV, $\langle\eta \rangle = f$, the potential
now has three phases: $\delta_4;~\delta_5$; and $\theta$ the phase
associated with the $\phi$ VEV.  There is now enough freedom in the theory so
that one linear combination of these phases gives rise to physical
CP-violating effects.  However having $\langle\eta\rangle \not= 0$ breaks
the $D$-discrete symmetry spontaneously, and domain walls will ensue
in the early Universe.  So, it is not clear this model makes any sense
cosmologically!

In general, however, if one introduces a sufficiently complicated Higgs
sector it is possible eventually to have some non-trivial CP-violating
phases.  A good example is Weinberg's 3 Higgs doublet model$~$\cite{Wein} in which there
are CP-violating phases associated to the coupling of the charged Higgs, $H^\pm$, to 
leptons and quarks.  Such models can give rise to new observable phenomena,
like the transverse muon polarization in $K_{\ell 3}$ decays.  One finds$~$\cite{trans}
\begin{equation}
\langle P_\perp^\mu\rangle \sim\frac{M^2_K}{M^2_H}~ 
{\rm Im}(g_{H\mu \nu}g^\star_{H d s}).
\end{equation}
This effect is interesting since, in principle, it can be larger than
the induced polarization from final state interactions  in $K^+\to \mu^+\pi^o\nu_\mu$
($\langle P_{\perp}^\mu\rangle_{FSI} \sim 10^{-6})~$ \cite{Z}, and is an effect which is not present in the CKM model.

\section{Supersymmetry and CP Violation}

It is difficult to take CP-violating phenomena produced by multi Higgs models
seriously, since there is no particular physical motivation for these models.
In this respect, supersymmetric (SUSY) extensions of the SM have a much better
pedigree.  Unfortunately, CP violation in these models is largely a function
of the {\bf assumed} pattern of the SUSY-breaking terms!  So these models,
at this stage, are not predictive.  In fact, as we shall see, there are
difficulties in SUSY models both with CP violation and flavor conservation,
which need careful control.

As is well known, the supersymmetric extension of the SM (SUSY SM) naturally
requires the appearance of 2 Higgs doublets $\chi$ and $\phi$.  The pure
Higgs potential in the SUSY SM does not contain either the $V_2$ or $V_3$ terms, so that $V = V_1$, with the various parameters
in $V_1$ taking particular values (related to the $SU(2)\times U(1)$ coupling
constants).  Without SUSY breaking, however, not only does the potential $V$
conserve CP but, in addition, it does not break $SU(2)\times U(1)$, since $\mu_1=\mu_2=0$!
As a result the introduction of soft SUSY breaking is a necessity to render these models realistic, not only by producing a splitting of particles and
sparticles in mass, but also to permit the breakdown of the electroweak
theory.

In the simplest scheme of SUSY breaking,\cite{sugra} this breaking is
mediated by gravitational strength forces and is {\bf assumed} to be
flavor blind.  Even in this very restricted context, CP-violating phases 
appear in four different places:
\begin{description}
\item{i)} In a complex gluino mass term:  $m_{1/2}\lambda_i\lambda_i$
\item{ii)} In the complex coefficient $A$ multiplying the Yukawa interactions
in the scalar sector: $A\Gamma_u\tilde Q_L\phi\tilde u_R +
A\Gamma_d\tilde Q_L\chi\tilde d_R$ + h.c.
\item{iii)} In the complex coefficient $B$ multiplying bilinear scalar terms:\\
$B\mu(\chi^TC\phi)$ + h.c.
\item{iv)} In the complex coefficient $\mu$ characterizing the SUSY conserving
Higgsino mass term: $\mu(\tilde\chi^TC\tilde\phi)$
\end{description}
One can check, however, that only 2 out of these possible 4 phases give rise
to physical effects.  For instance, the term $B\mu$,  is equivalent  to $\mu^2_{12}e^{i\delta_{12}}$ in the notation of Eq. (12). However, as we remarked
earlier, this phase {\bf by itself} does not give rise to physical 
CP-violating effects.

Nevertheless, as the simple example above demonstrates, in general once SUSY
breaking is introduced it is quite easy to get CP violation in the theory.
The difficulty really is the opposite, not in generating CP-violating
interactions but keeping the SUSY-breaking CP-violating contributions
from violating experimental constraints!  There are, in fact, two general
types of constraints in SUSY extensions of the SM that one must be careful to
satisfy.  The first of these deals specifically with CP violating phenomena
which preserve flavor, like the neutron electric dipole moment.  Already nearly
20 years ago, Dugan, Grinstein, and Hall \cite{DGH} remarked that the neutron
edm put powerful constraints on the CP-breaking phases appearing in SUSY
extensions of the SM.  They found, typically, that the neutron edm was given
by the formula
\begin{equation}
d_n \sim\left[300\left[\frac{100~{\rm GeV}}{\tilde m}\right]^2~\sin\phi_{A,B}
\right]~6.3\times 10^{-26}~{\rm ecm}.
\end{equation}
Here $\tilde m$ is a typical spartner mass and $\phi_{A,B}$ are the two independent CP-violating phases entering in the simplest SUSY breaking
structure described above.  To satisfy the present experimental bound on
$d_n$,\cite{PDG} one sees that these phases have to be of $O(10^{-3})$ for spartner masses 
$\tilde m\sim 100~{\rm GeV}$.

The second constraint on the flavor and CP structure of the SUSY breaking
terms arises from the appearance of FCNC interactions due to SUSY matter
entering at the loop level.  These loop contributions, unless appropriately
controlled, can lead to very large flavor changing effects, in complete
contradiction with experiment.  This whole subject has been analyzed in great
detail by many authors and it would take me too far afield to review it here.
For our more restricted purposes, connected with CP violation, a recent paper
of Dine, Kramer, Nir, and Shadmi \cite{DKNS} summarizes the relevant constraints
very effectively.  

The resulting structure of the SUSY CP violating effects
in the flavor sector depends, in general, on how one assumes that the SUSY
induced flavor violating effects are controlled.  Three mechanisms are
effective: \cite{DKNS}
\begin{description}
\item{i)} FCNC processes are suppressed through the imposition of near
{\bf universality} of the squark masses---$\Delta\tilde m^2 \ll \tilde m^2$
\item{ii)} FCNC processes can also be suppressed by dynamical
{\bf alignment} of the squark-quark couplings to the gluinos---$g_{\tilde gij}
\sim\delta_{ij}$
\item{iii)} Alternatively all loop processes can be suppressed by having {\bf heavy}
squarks and gluinos---$\tilde m\gg~{\rm TeV}$.
\end{description}

It turns out that,\cite{DKNS} in general, only in the case of weak
alignment one finds measurable SUSY breaking induced CP-violating effects.
Even then, the results are rather model dependent.  For example, about a year
ago, Masiero, Piai, and Vives \cite{MPV} constructed a model in which
$\delta_{\rm CKM} \equiv 0$. However, in their model CP violation due to SUSY breaking effects
gave $\epsilon\sim 10^{-3}$ (reflecting a phase in the squark-quark-gluino
mixing).  This model also gave a reasonable estimate for $\epsilon^\prime/\epsilon\sim 10^{-3}$, but predicted a small CP asymmetry for $B\to\psi K_S$.  This latter prediction, given the present data, makes this model
not viable.

Because all the experimental data on CP violation is in 
excellent agreement with the CKM model, at the moment for the SUSY SM all that one has
are constraints on squark mass splittings and on mixing in the gluino
couplings.  The typical parameters bounded are the quantities \cite{Abel}
\begin{equation}
\Delta_{ij} = \left(\frac{\tilde m_1^2-\tilde m_j^2}{\tilde m^2}\right)
g_{\tilde gij}
\end{equation}
for squarks associated with quarks of different helicities (L or R).  Typical
results of a recent analysis \cite{Becirevic} allow values in the B-sector (for either
helicity) as large as
\begin{eqnarray}
{\rm Re}~\Delta_{13} &\cong& 2\times 10^{-2} \nonumber \\
{\rm Im}~\Delta_{13} &\cong& 10^{-2},
\end{eqnarray}
without giving noticeable effects in the data.

\section{What Should We Look For?}

It is clearly of fundamental importance to pin down the unitarity
triangle, to check whether or not all {\bf flavor changing} CP
violation effects  originate solely from the single CKM phase $\delta_{\rm CKM}$.
As remarked by Buras \cite{Buras} at this meeting, in the context of the CKM model,
measurements of the (CP-conserving) sides of the unitarity triangle are as important
as the measurement of an explicit CP-violating angle.  In a more general context, however,
what is crucial is to ascertain whether there are any additiional CP-violating
phases, besides $\delta_{\rm CKM}$.

In this respect, perhaps the most clear signals of CP-violation occur
in extended Higgs models, where CP violating effects
manifest themselves by the appearance of Higgs bosons with mixed CP
properties.  As an example, it is well known that the three neutral scalars
$h,~H$ and $A$ in the SUSY SM have well defined CP properties, with the first two fields being $0^+$ objects and $A$ being a $0^-$ excitation.  This,
however, may no longer hold at loop order, since CP-violating loop effects
involving stops and sbottoms can mix $A$ with $h$ and $H$.\cite{Pilaftsis}

In general, therefore, it is important to look for the presence of possible CP-violating couplings of Higgs bosons.  CP violation, for example, allows
the lightest Higgs boson $h$ to couple to two photons both through an
$F^2$ and an $F\tilde F$ term:
\begin{equation}
{\cal{L}} = \frac{\alpha}{\pi}[aF^{\mu\nu}F_{\mu\nu} + bF^{\mu\nu}
\tilde F_{\mu\nu}]h.
\end{equation}
In practice, however, it is going to be quite difficult to measure possible
CP-violating coefficients (like the ratio $b/a$ above).  At the LHC,
typically, one is sensitive to CP-odd mixing at the level of about
30\%.\cite{Conwayetal}  In this respect, an NLC would be much more effective
at detecting Higgs sector CP violation, with the sensitivity to CP-odd mixings
being pushed to about the 4\% level. \cite{Conwayetal}

It is natural to ask if there are other areas, besides the Higgs sector, where
one should look for hints regarding CP violation.  In this context, it is
important to note that we know from the existence of a baryon asymmetry in the
Universe that there must be other CP-violating phases besides $\delta_{\rm CKM}$,
since the SM does not seem to be able to produce such an asymmetry.\cite{antiShap}  It is barely possible that this asymmetry
might be produced at the electroweak phase transition in the SUSY SM. However,  the
parameter space for this to happen is extremely restricted.  Basically, as
discussed extensively by Quiros and collaborators,\cite{Quiros} it is crucial to have a
sufficiently strong first order transition to prevent the baryon asymmetry produced at the electroweak phase transition from being erased. For this to obtain in the SUSY SM one needs a
very light stop ($\tilde{m}_t\sim 140$ GeV) and the lightest Higgs boson must have
a mass near the edge of discovery ($m_h \sim 115$ GeV).  So, although probably
not totally ruled out yet, this possibility seems highly unlikely to me.

A much more intriguing possibility, and one which I believe is much more likely, is that the CP phase
responsible for the baryon asymmetry in the Universe is actually connected to
CP-violating phases in the neutrino sector.  The scenario for this to happen is well known and was first outlined
more than 15 years ago by Fukugita and Yanagida.\cite{FY}  What these authors 
observed was that, in theories with heavy Majorana neutrinos, the out of
equilibrium decay of these heavy neutrinos (with masses $M\geq 
10^{10}$ GeV) can establish a lepton asymmetry in the early Universe at
temperatures of order $T\sim M$.  Such a lepton asymmetry would naturally give a baryon asymmetry, through
the Kuzmin Rubakov Shaposhnikov (KRS) mechanism. \cite{KRS}  What KRS showed is that B+L-violating interactions are expected to be in equilibrium for a large temperature range (100 GeV $\leq T \leq 10^{12}$ GeV). Thus, as the 
Universe cools below $T\sim M$ these processes serve to erase any created (B+L)-asymmetry, and effectively serve to trasmute any lepton asymmetry into a corresponding baryon asymmetry.

Although this is an attractive scenario for generating the Universe's
baryon asymmetry, as discussed by Hambye \cite{H}
and Hernandez \cite{He} at this meeting it is, in general, difficult to relate directly the
CP-violating phases at the root of the original lepton asymmetry with possible low energy
CP-violating phenomena in the neutrino sector.  Nevertheless, it is interesting
to examine how one can conceivably extract from experiment information on
CP violation of neutrinos.

Evidence for CP violation in the neutrino sector can come from two places:
neutrino oscillations and neutrinoless double beta decay.  These two processes
are, in some sense, complementary \cite{Barger} since they are sensitive to
different CP-violating phases.  Measuring the difference in oscillations between
neutrinos and antineutrinos gives direct evidence for CP violation, while the
rate for neutrinoless double beta decay provides more indirect information
on CP violation.  As Blondel \cite{Blondel} emphasised at this meeting, in either
case the experimental challenges are enormous!

Neutrino (and antineutrino) oscillations can
give information on $\delta^\ell_{\rm CKM}$, the leptonic equivalent of the
CKM CP violating phase.  Neutrinoless double beta decay, on the other hand,
is most sensitive to a CP violating phase $\varphi_M$ which enter only if neutrinos are Majorana particles. Let me briefly examine these two processes in turn.

For oscillation experiments to successfully detect CP-violating effects in
the neutrino sector the angle $\theta_{13}$ in the leptonic mixing matrix must
be near the CHOOZ bound [$\sin^22\theta_{13} < 0.1$ \cite{CHOOZ}]. In fact, it is easy to check that, as this
angle tends to zero, the difference between the oscillation probabilities
of neutrinos and antineutrinos disappears.  For example, for $\nu_\mu$ to $\nu_e$
transitions, one finds
\begin{eqnarray}
P[\nu_\mu\to\nu_e] &=& a+b\sin 2\theta_{13}\sin\delta^\ell_{\rm CKM} 
\nonumber \\
P[\bar\nu_\mu\to\bar\nu_e] &=& a-b\sin 2\theta_{13}\sin\delta^\ell_{\rm CKM}.
\end{eqnarray}
Thus, the accuracy with which one can determine $\sin\delta^\ell_{\rm CKM}$ is
directly related to the magnitude of $\theta_{13}$.

For neutrinoless double beta decay, \cite{Petcov} the situation is different, since one can detect the presence of a CP-violating Majorana
phase $\varphi_M$ even in the limit when $\theta_{13}$ vanishes. 
Indeed, neglecting $\theta_{13}$ altogether yields for the effective
neutrino mass $M_{ee}$ measured in this process a very simple formula. For the case of a normal hierarchy where
$m_3 \gg m_2 \simeq m_1$, with $m_2 \simeq m_1 = m_e$ the mass of the
neutrino  measurable, in principle, in Tritium beta decay, one has
\begin{equation}
M_{ee} = m_e \left|\cos^2\theta_{12} + e^{i\varphi_M}\sin^2\theta_{12}\right|.
\end{equation}
It is an open question whether even with the observation of neutrinoless
double beta decay Eq. (23) will allow an extraction of $\varphi_M$, given the large
theoretical error which accompanies the extraction of $M_{ee}$ from
experiment.\cite{Barger} \cite{Petcov}

Unfortunately, as alluded to earlier, neither $\varphi_M$ or $\delta^\ell_{\rm CKM}$ are directly related to the CP-violating phases that control leptogenesis, except in very particular circumstances.  The baryon asymmetry is directly
related to the lepton asymmetry produced at temperatures of order
$T\sim M$.  In turn, the lepton asymmetry is proportional to the
CP asymmetry in the decay of the heavy neutrino.  One has, for the SM,\cite{Berezhiani}
\begin{equation}
\eta_B = -\frac{8}{15} \eta_L = -\frac{8}{15}
\left[\frac{\kappa}{g*}\right]\epsilon.
\end{equation}
Here $\kappa$ is a washout factor connected with how the heavy neutrino
decays go out of equilibrium,\cite{FY} while $g*\sim 100$ is the number of
degrees of freedom at temperatures of $O(T\sim M)$.  The CP asymmetry
factor $\epsilon$ in Eq. (24) depends in detail on the phase structure
of the heavy neutrino Yukawa coupling $h_\nu$ coupling the leptonic 
doublet $L_L$ to $N_R$ and the Higgs doublet.  The phases, $\delta^\ell_{\rm CKM}$ and
$\varphi^M$, on the other hand, also depend on the phase structure of the 
electron Yukawa coupling $h_\ell$ coupling the leptonic doublet $L_L$
to $\ell_R$ and the Higgs doublet.

\section{Concluding Remarks}

Obviously, at this stage, the most important task still in front of us is to
get {\bf additional} experimental information on CP violation.  Fortunately,
the prospects for doing so are very good, since there are many experimental
efforts precisely targeted in this direction.  These range from experiments
presently underway at the B factories at SLAC and KEK, with the BaBar and
Belle detectors, to dedicated collider B-experiments planned for the
Tevatron and the LHC.\footnote{One should not forget the important role
that CDF and $\rm D^0$ can play in the near future at the Tevatron to elucidate aspects of B physics not accessible to the B factories.}
The list of experiments does not end here.  Powerful information can come also
from searches for $K_L\to\pi^o\nu\bar\nu$ \cite{T} (and its charged counterpart
$K^+\to \pi^+\nu\bar\nu$, already seen by the Brookhaven experiment
E787 \cite{Bry}), and new experiments to detect the neutron and electron
electric dipole moment with an order of magnitude or more precision. \cite{edm}
Finally, searches for CP violation in the neutrino sector are expected to be undertaken as the new generation of high flux neutrino experiments get underway. \cite{Blondel}

One of the principal goals of this experimental program is to understand if
the simple CKM paradigm explains {\bf all} CP violating phenomena in the
hadronic sector, or if there are other CP-violating phases.  Of equal
importance is to see if there is any signal of CP violation in the leptonic
sector- something one is led to expect by analogy to the hadronic sector.

On the theoretical side, in my view, it is important to take seriously some of the
hints we have from experiment regarding CP violation.  Recall, from our discussion,
that CP is a good symmetry in 10 dimensions, but is broken in our $d=4$
world.  Furthermore, even in four dimensions it is difficult to break CP.
There is, apparently, no $\theta F\tilde F$ term and there is at least indirect evidence that CP violation originates from the presence of scalars. For the SM this leads to the CKM paradigm. However, even with a more complicated scalar sector than that
is the SM, the fact that there are no FCNC argues for the absence of terms
which could introduce more CP phases.  In SUSY theories, the problem is not how
to generate CP violation but rather how to organize the SUSY breaking so as not
to generate too large CP-violating effects.

To all of these restrictions, one must add the fact that the observed CP violating phenomena
have  quite different magnitudes [$\epsilon\sim 10^{-3}$, $\epsilon^\prime/\epsilon\sim 10^{-3}$; $a_{B\to\psi K_S}\sim 1$]. However, these disparate meausurements all appear to be
explained by the same (large)  CP-violating phase $\delta_{\rm CKM}$.  This leads one to speculate, as was done by Abel \cite{Abel} in this meeting, that perhaps {\bf all} CP violation originates from the presence of a single geometrical phase, \cite{AO}
associated with the compactification from $d=10$ to $d=4$.  If I had to make a
guess, I would imagine that this Ur-CP phase $\delta_o$ is related to some
topological invariant of the associated geometry.  An appealing formula is
simply
\begin{equation}
\delta_o = \frac{2\pi}{N_g}
\end{equation}
with $N_g$ being the number of generations.\footnote{This formula gives no
CP-violation in the case of two generations. This is not necessarily the
case if neutrinos are Majorana particles, although the presence of such Majorana phases is not mandated.}
It is amusing to note that the CKM phase is near $60^\circ$, in the usual
convention.\cite{Kleinknecht}

\section*{Acknowledgements}

I am grateful to Xinmin Zhang and Tao Huang for discussions on some of the material presented here. This work has been supported in part by the Department of Energy under Contract No. FG03-91ER40662, Task C.

\end{document}